\documentclass[fleqn,usenatbib]{mnras}

\usepackage{newtxtext,newtxmath}

\usepackage[T1]{fontenc}

\DeclareRobustCommand{\VAN}[3]{#2}
\let\VANthebibliography\thebibliography
\def\thebibliography{\DeclareRobustCommand{\VAN}[3]{##3}\VANthebibliography}

\usepackage{graphicx}	
\usepackage{amsmath}	
\usepackage{subcaption}
\usepackage{lineno}
\usepackage{graphicx}

\usepackage{soul}
\usepackage{amsmath}
\usepackage{xspace}
\usepackage{xifthen}
\usepackage{eso-pic}

\definecolor{forestgreen}{HTML}{228B22}
\definecolor{urlblue}{HTML}{000000}

\mathchardef\mhyphen="2D

\newlength{\dhatheight}

\newcommand{\bandvar}[2][]{%
  \ifthenelse{\isempty{#1}}{\var{#2}}{\var{#2\_#1}}%
}

\newcommand{\var}[1]{\ensuremath{\texttt{\MakeUppercase{#1}}}\xspace}

\providecommand\physrep{\ref@jnl{Phys.~Rep.}}%
\providecommand\apjs{\ref@jnl{ApJS}}%
\providecommand{\jcap}{\ref@jnl{JCAP}}%

\input{hyperlink-year-only-natbib-patch}
\defcitealias{Riley:2025}{Paper I}
\defcitealias{Shipp:2025}{Paper II}

\graphicspath{{./figures/}{./}}

\title[The MZR of disrupting satellites]{Auriga Streams III: the mass-metallicity relation does not rule out tidal mass loss in Local Group satellites}

\author[A.H.~Riley et al.]{Alexander H.~Riley,$^{1,2}$\thanks{E-mail: alexander.riley@fysik.lu.se}
Rebekka Bieri,$^{3}$
Alis J.~Deason,$^{1}$
Nora Shipp,$^{4}$
Christine M.~Simpson,$^{5}$ \newauthor
Francesca Fragkoudi,$^{1}$
Facundo A.~G\'{o}mez,$^{6}$
Robert J.~J.~Grand,$^{7}$
and Federico Marinacci$^{8,9}$
\\
$^{1}$Institute for Computational Cosmology, Department of Physics, Durham University, South Road, Durham DH1 3LE, UK\\
$^{2}$Lund Observatory, Division of Astrophysics, Department of Physics, Lund University, SE-221 00 Lund, Sweden\\
$^{3}$Department of Astrophysics, University of Zurich, 8057 Zurich, Switzerland\\
$^{4}$Department of Astronomy, University of Washington, Seattle, WA 98195, USA\\
$^{5}$Argonne Leadership Computing Facility, Argonne National Laboratory, Lemont, IL 60439, USA\\
$^{6}$Departamento de Astronom\'{i}a, Universidad de La Serena, Av.~Juan Cisternas 1200 Norte, La Serena, Chile\\
$^{7}$Astrophysics Research Institute, Liverpool John Moores University, 146 Brownlow Hill, Liverpool, L3 5RF, UK\\
$^{8}$Department of Physics \& Astronomy `Augusto Righi', University of Bologna, via Gobetti 93/2, I-40129 Bologna, Italy\\
$^{9}$INAF, Astrophysics and Space Science Observatory Bologna, Via P. Gobetti 93/3, I-40129 Bologna, Italy
}

\date{Accepted 2025 December 16. Received 2025 December 15; in original form 2025 September 11}

\pubyear{2025}

\begin{document}
\label{firstpage}
\pagerange{\pageref{firstpage}--\pageref{lastpage}}
\maketitle


\begin{abstract}
    The mass-metallicity relation is a fundamental galaxy scaling law that has been extended to the faintest systems in the Local Group.
    We show that the small scatter in this relation, which has been used to argue against tidal mass loss in Local Group satellites, is consistent with the level of disruption in the Auriga simulations.
    For every accreted system in Auriga, we compute stellar masses and metallicities two ways: considering the total system (bound + lost material) and only considering the progenitor.
    Accreted systems in Auriga have a tight relation between total stellar mass and metallicity, with scatter at a fixed stellar mass driven by age.
    When only considering the progenitor, the tidally evolved mass-metallicity relation has similar scatter ($\sim$0.27 dex) as observed for the Local Group satellites ($\sim$0.23 dex).
    Satellites that lie above the evolved relation have experienced substantial mass loss and typically have low metallicity for their total stellar mass.
    Even satellites that fall exactly on the evolved relation can lose over half of their stellar mass.
    Only satellites substantially below the evolved relation are reliably intact.
    Based on their offset from the observed relation, we predict which Milky Way and M31 satellites have tidal tails waiting to be discovered.
\end{abstract}



\section{Introduction}

The mass-metallicity relation is a fundamental scaling law connecting the stellar mass of a galaxy to its metal contents \citep[e.g.][]{Tremonti:2004, Kewley:2008, Mannucci:2010, Maiolino:2019}.
The relation is governed by the complex interplay between star formation, stellar evolution, accretion, and baryonic feedback.
It has been extended to the faintest galaxies within the Local Volume \citep{Simon:2007, Kirby:2008, Kirby:2013, Kirby:2020}, many of which are satellites of the Milky Way and M31.

The observed scatter\footnote{We adopt the RMSE, $\sigma_\text{scatter} \equiv \left[ \sum \left( \text{[Fe/H]}_\text{obs} - \text{[Fe/H]}_\text{model}{} \right)^2 / N \right]^{1/2}$.} in this relation is remarkably small, $\sim$0.15 to 0.25~dex depending on adopted measurements.
It is conventionally argued \citep[e.g.][]{Simon:2007, Simon:2019} that this small scatter indicates that the Local Group satellites have not experienced substantial mass loss.
This is because individual systems would move toward smaller mass and higher mean metallicity if the system begins with a negative radial metallicity gradient and low-metallicity stars in the outskirts are preferentially lost.
This would tend to increase scatter in the correlation.

Many cosmological hydrodynamical zoom-in simulations of Milky Way-mass galaxies claim to match the mass-metallicity relation \citep{Genina:2019, Grand:2021, Applebaum:2021}, often with similar or smaller scatter than the observations.
Recent analyses of such simulations \citep{Shipp:2023, Shipp:2025, Riley:2025} show that their satellites have experienced much more tidal disruption than previously understood.
How do these simulations produce mass-metallicity relations with small scatter, despite such a high level of mass loss?
Is the observed mass-metallicity relation indicative of an intrinsic relation, or possibly one that has evolved due to tides?

Our goal in this work is to examine these questions using the Auriga simulations.
We restrict our analysis to satellites that reach $M_\star \gtrsim 5\times10^5$~M$_\odot$, excluding the ultrafaint systems that may not follow the same mass-metallicity relation \citep{Fu:2023} due to higher sensitivity to supernova feedback strength and other galaxy formation physics \citep{Agertz:2020}.

\section{Disrupting satellites in Auriga} \label{sec:data}

We use the suite of Milky Way-mass haloes from the Auriga project \citep{Grand:2017, Grand:2024}.
These haloes were selected from the EAGLE dark-matter-only 100 Mpc box \citep{Schaye:2015} to have $M_\text{200c} = 1 - 2 \times 10^{12}$~M$_\odot$ and satisfy an isolation criterion.
They were resimulated with the moving-mesh code \textsc{Arepo} \citep{Springel:2010, Pakmor:2016} and a galaxy formation model detailed in \citet{Grand:2017, Grand:2024}.
The Auriga model produces spiral disc galaxies that are broadly consistent with a number of observations including stellar masses, sizes, and rotation curves \citep{Grand:2017}, H~\textsc{i} gas distributions \citep{Marinacci:2017}, stellar disc warps \citep{Gomez:2017}, stellar bars \citep{Fragkoudi:2020, Fragkoudi:2025}, satellite galaxies \citep{Simpson:2018}, stellar haloes \citep{Monachesi:2019}, and magnetic fields \citep{Pakmor:2017}.
The simulations are described in further detail in \citet{Grand:2017, Grand:2024}.

\citet{Riley:2025} and \citet{Shipp:2025}, hereafter \citetalias{Riley:2025} and \citetalias{Shipp:2025}, presented a uniform and complete catalogue of all systems that accreted onto the Milky Way-mass hosts in Auriga.
In addition to identifying the star particles associated with each system, they also assessed which are bound to the progenitor at present day according to SUBFIND \citep{Springel:2001}.
They analysed three different numerical resolutions available in Auriga; we focus here on the six haloes that were simulated at `level~3' resolution with baryonic element (dark matter particle) masses of 6.7(36)$\times 10^3$~M$_\odot$ and a minimum softening length of 188~pc.
This resolution is sufficient to study accreted systems down to $M_\star$~$\sim$~$5\times10^{5}$~M$_\odot$ with over 100 star particles. 


The Auriga simulations trace individual abundances for nine elements: H, He, C, O, N, Ne, Mg, Si, and Fe.
The abundances of individual star particles are inherited from the gas cell they are born from.
We define the metallicity of a star particle as
\begin{equation} \label{eqn:feh-definition}
    \text{[Fe/H]} = \log_{10} \left( \frac{M_\text{Fe}}{M_\text{H}} \right) - \log_{10} \left( \frac{55.845}{1.008} \right) - (7.46 - 12),
\end{equation}
where $M_X$ is the ratio of mass in species $X$ to the total mass of the parent gas cell, the second term converts mass to number density using the atomic mass of each species, and the third term sets $\text{[Fe/H]}_\odot \equiv 0$ \citep{Asplund:2021}.
When computing the mean metallicities for an accretion event, we follow conventions for resolved star observations in the Local Group.
In particular, we weight each star particle by its present-day stellar mass and compute the mean of [Fe/H] as defined in Equation~\ref{eqn:feh-definition} \citep[see Section 3.1 of][]{Escala:2018}.
For each system\footnote{We remove 13 systems (of 419 total) that are phase-mixed with intrinsic $M_\star \lesssim 10^7$~M$_\odot$ and [Fe/H] $\gtrsim$ -0.9. These are contaminated by merger-driven starbursts, becoming outliers in mass-metallicity-age.} we compute the mean metallicity two ways: across all stars formed in that system (`total') and only those bound to the progenitor (`bound').
We refer to the resulting mass-metallicity relations as `intrinsic' and `(tidally) evolved', respectively.
We denote the fraction of total stellar mass that remains bound to the progenitor as $f_\text{bound}$\footnote{We refer to Sections 1 and 3 of \citetalias{Riley:2025} for adopted nomenclature and definitions (e.g., `bound', `progenitor', `total stellar mass' of a system).}.

We caution that the Auriga model produces central galaxies, satellites, and stellar haloes that are more metal rich than in observations \citep{Monachesi:2019, Grand:2021, Kizhuprakkat:2024}.
We focus on the evolution of the mass-metallicity relation as Auriga satellites lose stellar mass, which depends on \textit{relative} changes.
In Section \ref{sec:local-group-sats} we compare the offsets from the mass-metallicity relation for both Auriga and observations, but only after normalizing by the scatter about their respective relations, which also mitigates this issue.
Similarly, if satellites in Auriga disrupt too readily (either from numerical resolution or unrealistic physical properties), the general qualitative trends reported in this work should still hold.

\section{Intrinsic vs.~evolved mass-metallicity relation} \label{sec:relations}

\begin{figure}
    \centering
    \includegraphics[width=1.0\linewidth]{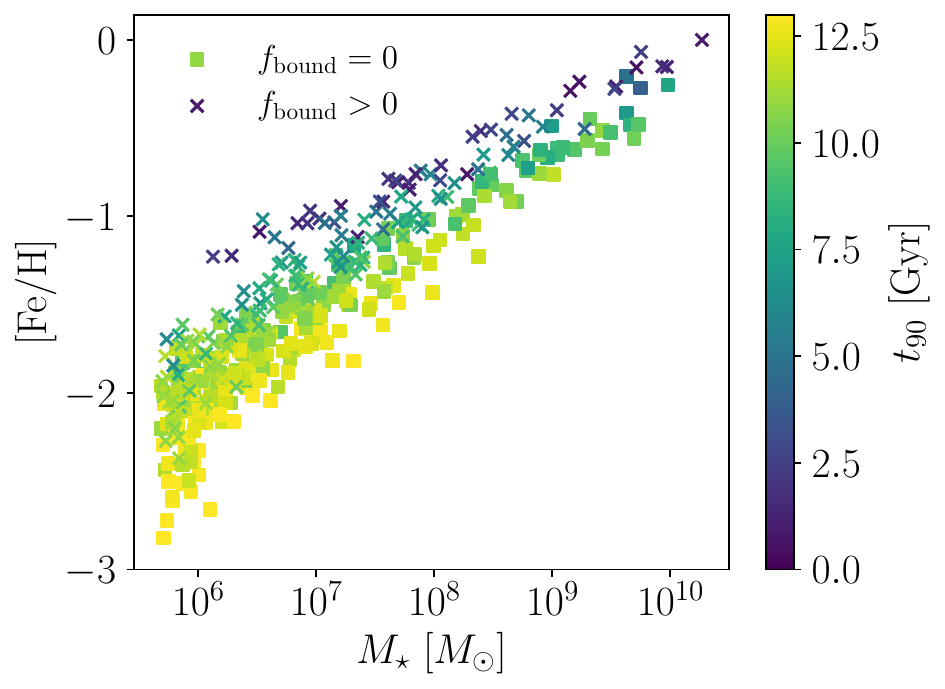}
    \caption{The intrinsic mass-metallicity relation for accreted systems in Auriga.
    Squares indicate systems that have been fully disrupted ($f_\text{bound} = 0$) while crosses indicate systems that still have a bound progenitor at the present day ($f_\text{bound} > 0$).
    The scatter at fixed total stellar mass is driven by age (colour of points), parametrised here as $t_{90}$.}
    \label{fig:mass-feh-t90}
\end{figure}

In Figure~\ref{fig:mass-feh-t90} we present the intrinsic mass-metallicity relation for all accreted systems in Auriga, regardless of whether or not they have a bound progenitor at the present day.
The scatter at fixed total stellar mass is driven by age\footnote{Parametrised as the lookback time by which 90~per~cent of stellar mass formed and denoted as $t_{90}$.}; older objects at a given stellar mass have lower metallicity and are more likely to be fully disrupted.
This naturally arises from a mass-metallicity relation that evolves with redshift due to self enrichment \citep[e.g.][]{Ma:2016, Torrey:2019}, combined with environmental quenching and tidal disruption experienced by objects from a cosmological accretion history.
We also find that fully disrupted systems have higher [$\alpha$/Fe] than those that survive to the present day, a familiar result in both simulations \citep{Robertson:2005, Font:2006, Fattahi:2020, Grimozzi:2024, Pathak:2025} and observations \citep{Ji:2020, Hasselquist:2021, Naidu:2022} that is driven by the same physical processes.


\begin{figure*}
    \centering
    \includegraphics[width=1.0\linewidth]{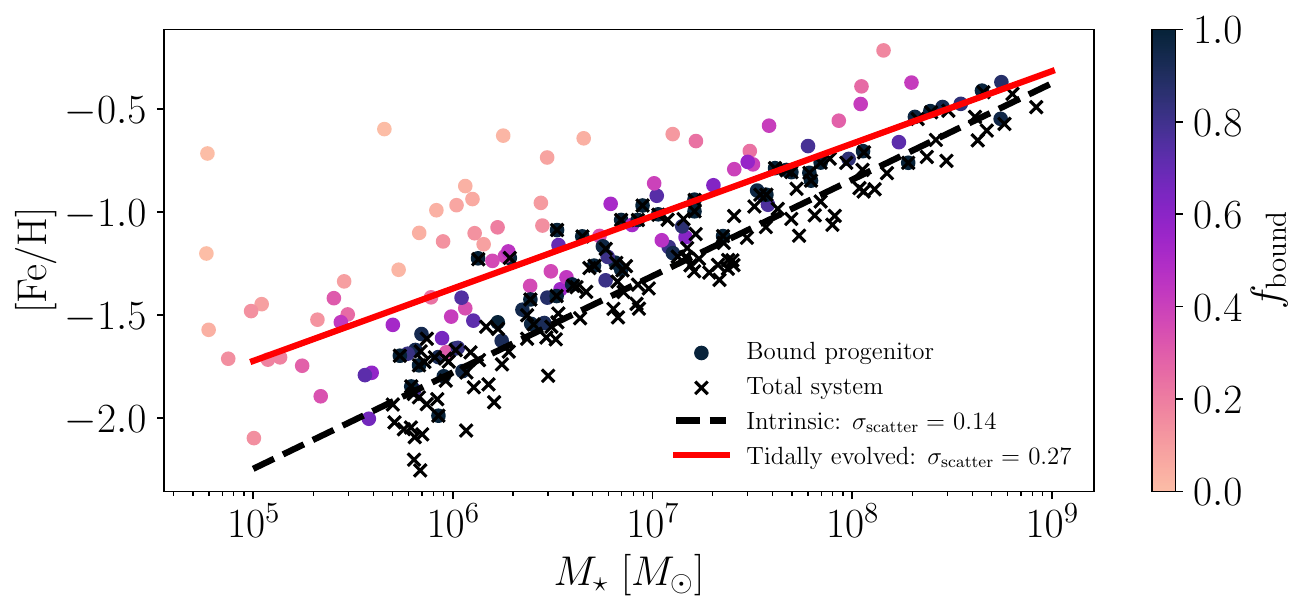}
    \caption{The intrinsic and tidally evolved mass-metallicity relations for Auriga satellites that have $f_\text{bound} > 0$.
    Individual systems are shown both considering bound progenitors (circles) and the total system (crosses).
    We provide linear fits to both bound (solid red) and total (dashed black) cases, along with the vertical scatter about these relations.
    We only include systems that have 10 or more bound star particles at the present day.
    }
    \label{fig:mass-feh-fbound}
\end{figure*}

In Figure~\ref{fig:mass-feh-fbound} we present both the intrinsic and evolved mass-metallicity relations for surviving Auriga satellites (i.e.~$f_\text{bound} > 0$).
Every satellite is represented in the Figure twice: crosses indicate values when considering bound and lost material (at the same locations as in Figure~\ref{fig:mass-feh-t90}), while the circles only consider stars that remain bound to the progenitor at the present day.
As satellites disrupt, they move towards lower (bound) stellar mass and higher mean metallicity.
The increase in mean metallicity is a result of negative radial metallicity gradients in Auriga\footnote{We discuss ignoring metallicity gradients in Appendix~\ref{app:ignore-gradients}.} \citep{Khoperskov:2023, Orkney:2023, Carrillo:2024} -- as lower metallicity stars in the outer regions are removed first, the remaining bound progenitor has a higher mean metallicity than the overall system.
We note that negative metallicity gradients are common in observed low-mass galaxies in the Local Group and beyond \citep{Kirby:2011, Leaman:2013, Taibi:2022, Barbosa:2025}.

We fit linear models to both the intrinsic and tidally evolved data for systems that have $f_\text{bound} > 0$.
Due to tidal mass loss and negative metallicity gradients, the evolved relation sits higher and has a larger scatter (0.27) than the intrinsic relation (0.14).
The tidally evolved relation, which only considers the bound progenitor, is the simulated analogue to the measurements reported in \citet{Kirby:2013} and similar analyses.

The scatter in the tidally evolved relation is primarily driven by mass loss.
At fixed bound stellar mass, satellites with high $f_\text{bound}$ are generally below the tidally evolved relation, while satellites with low $f_\text{bound}$ are above this relation.
Even satellites that fall exactly on the relation can lose substantial amounts of stellar mass.
We also note that individual points `invert' their relationship to the mass-metallicity relation.
Satellites that have low $f_\text{bound}$ move in the mass-metallicity plane from being below the intrinsic relation to being above the tidally evolved relation.
Satellites that have high $f_\text{bound}$ begin above the intrinsic relation and remain in the same location of the mass-metallicity plane, but the relation shifts such that they end up below the tidally evolved relation.
There is a modest decrease in scatter about the tidally evolved relation with increasing stellar mass, while the scatter is more uniform for the intrinsic relation.

Finally, we note that there is a collection of satellites with $f_\text{bound} \simeq 1$ that fall on or slightly above the tidally evolved relation.
These systems are star forming, recently accreted satellites that have the highest possible metallicity for their total stellar mass (see Figure~\ref{fig:mass-feh-t90}).
The agreement in mass-metallicity relations between different morphological classes of observed galaxies \citep{Kirby:2013} and between satellite and field galaxies in some simulations \citep{Applebaum:2021} may stem from comparing the intrinsic relation of recently accreted (or isolated) star forming systems to the tidally evolved relation of quiescent systems.
We show these galaxies in Figure~2 but exclude them elsewhere in this work.

\section{The Local Group satellites} \label{sec:local-group-sats}

In Section \ref{sec:relations} we established a connection between how disrupted a satellite in Auriga is and where it sits in the mass-metallicity plane relative to the tidally evolved relation.
Now we connect this result to the observed satellites of the Milky Way and M31.
We consider satellites that are brighter than $M_V = -7.7$ \citep{Simon:2019} and have literature [Fe/H] measurements derived from spectroscopy.
We adopt values reported in \citet{Pace:2025} for all stellar mass measurements\footnote{In practice, \citet{Pace:2025} take literature $V$-band luminosities and adopt a uniform $M_\star/L_V = 2$.} and augment their spectroscopic measurements with values for high-mass satellites from either \citet{McConnachie:2012} or \citet{Kirby:2013}.
Specific measurements and their original sources are listed in Appendix~\ref{app:individual-sats}.

\begin{figure}
    \centering
    \includegraphics[width=1.0\linewidth]{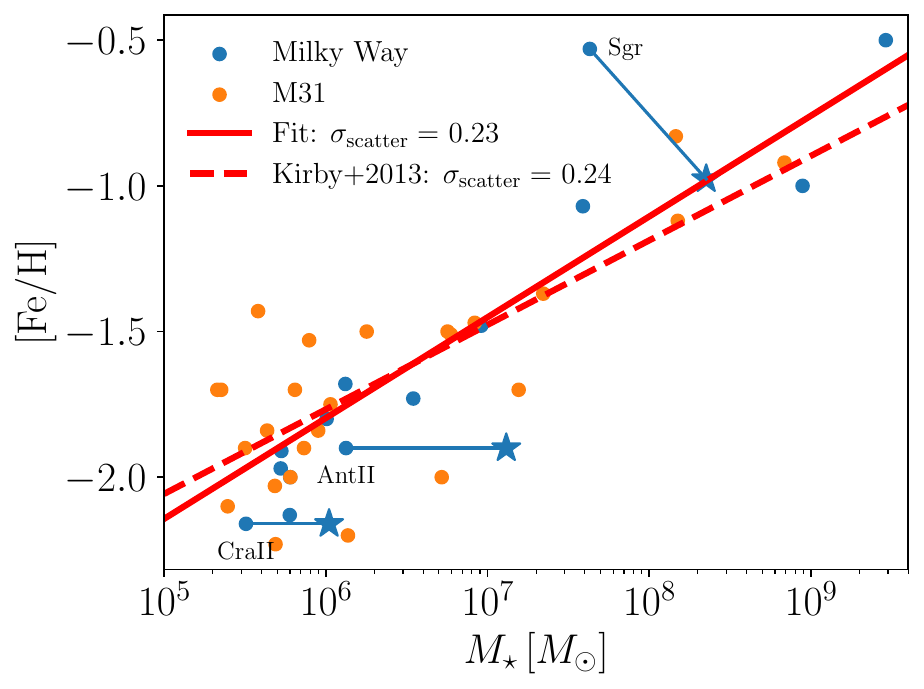}
    \caption{The mass-metallicity relation for observed satellites of the Milky Way (blue points) and M31 (orange points).
    We provide a linear fit to these satellites (solid) and the relation from \citet{Kirby:2013} (dashed).
    We also highlight individual satellites known or believed to be disrupting and their reconstructed `total' values (star symbols, see text for details).}
    \label{fig:mass-feh-observations}
\end{figure}

In Figure~\ref{fig:mass-feh-observations} we present these measurements along with the relation reported in \citet{Kirby:2013}\footnote{For consistency with our adopted stellar masses, we use the luminosity-metallicity relation reported in Equation~3 of \citet{Kirby:2013} and apply the same $M_\star/L_V = 2$.}.
We note that the fit reported in \citet{Kirby:2013} includes ultrafaint systems down to $10^3$~M$_\odot$, which do not appear to continue the relation and instead flatten at [Fe/H]~$\sim$~$-2.6$ \citep{Fu:2023}.
For a consistent treatment, we fit a new relation as done for the Auriga satellites (Figure~\ref{fig:mass-feh-fbound}).
This fit reduces the scatter from 0.24 to 0.23, which is comparable to the 0.27 for bound satellites in Auriga.
We emphasize that this is the first time a simulation has shown that a mass-metallicity relation with small scatter (comparable to observations) can still emerge from a system of satellites undergoing substantial tidal disruption.

\begin{figure}
    \centering
    \includegraphics[width=1.0\linewidth]{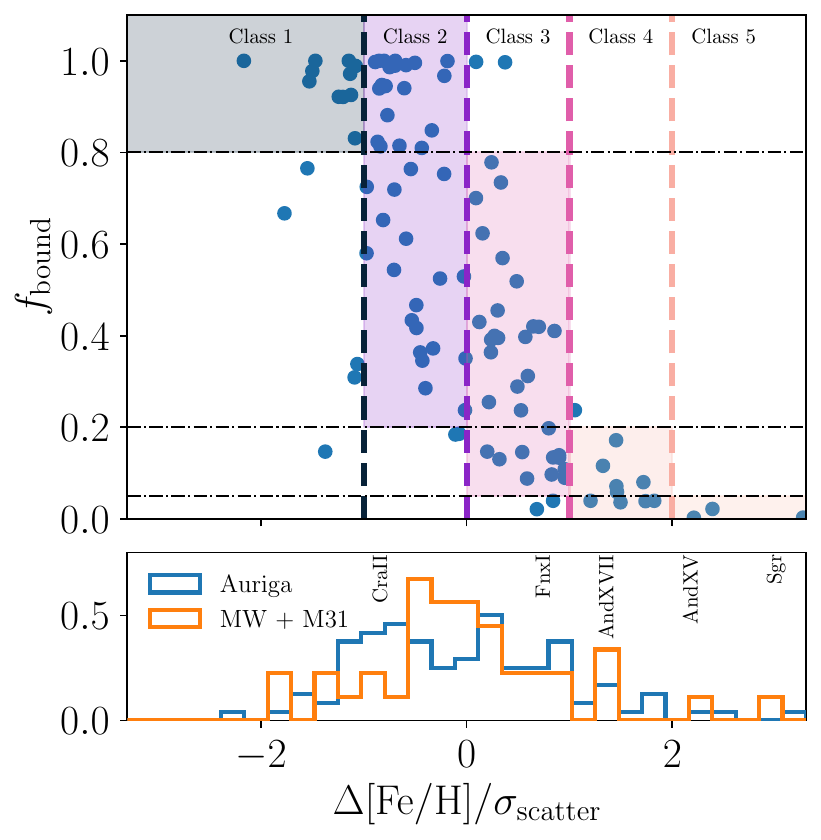}
    \caption{Top panel: $f_\text{bound}$ vs.~scaled offset from the evolved mass-metallicity relation for Auriga satellites.
    The dashed lines are the boundaries between the five broad categories of mass loss detailed in Section~\ref{sec:local-group-sats} and Appendix~\ref{app:individual-sats}.
    Bottom panel: histogram of scaled offsets for both Auriga and Milky Way and M31 satellites from their respective mass-metallicity relations.}
    \label{fig:feh-diff-fbound}
\end{figure}

We compute the vertical (i.e.~at fixed mass) offset from the fit $\Delta$[Fe/H] for each satellite, which is positive (negative) for satellites that sit above (below) the relation.
In the top panel of Figure~\ref{fig:feh-diff-fbound} we present these offsets and show they broadly correlate with $f_\text{bound}$ for the Auriga satellites, as seen in Figure~\ref{fig:mass-feh-fbound}.

We can leverage this connection between $f_\text{bound}$ and $\Delta$[Fe/H] to predict which Local Group satellites are actively disrupting.
This is detailed in Appendix~\ref{app:individual-sats}, but in brief, we assign a Class 1 through 5 based on the offset $\Delta$[Fe/H] normalised by the $\sigma_\text{scatter}$ about the relation.
The boundaries between the different classes are shown as dashed lines in Figure~\ref{fig:feh-diff-fbound} and largely capture different categories of mass loss in Auriga, from unlikely to be losing mass (Class 1) to guaranteed to have experienced extreme disruption (Class 5).
In particular, we highlight the following quiescent systems predicted to be disrupting at high confidence:
\begin{itemize}
    \item Class 3 (0.05\ <\ $f_\text{bound}$\ <\ 0.8): Fornax, Leo~II, NGC~147, And~I, II, III, VI, XIX, XXI, XXVIII, XXIX
    \item Class 4 ($f_\text{bound} < 0.2$): And~XVII, And~XVIII, Cassiopeia II
    \item Class 5 ($f_\text{bound} < 0.05$): Sagittarius, And~XV
\end{itemize}
The offset $\Delta$[Fe/H] and predicted disruption Class for individual Local Group satellites are reported in Appendix~\ref{app:individual-sats}.

It may appear surprising that so many Local Group satellites could be significantly disrupted, especially given current detections.
However, the tidal tails emanating from these systems are likely too faint to be detected in current resolved-star imaging, so the satellites would appear intact \citep{Shipp:2023}.
These low surface brightness features may be visible by upcoming facilities including Rubin~LSST, Euclid, and Roman.

To illustrate how far satellites can move in the mass-metallicity plane, in Figure~\ref{fig:mass-feh-observations} (star symbols) we provide reconstructions of the `total' system for three prominent examples of disrupting Milky Way satellites: Sagittarius, Antlia~II, and Crater~II.
For Sagittarius we adopt an $f_\text{bound}$~$=$~0.19 
\citep{Niederste-Ostholt:2010} and assume that the progenitor has mean [Fe/H]~$=$~$-0.58$, while the tails are more metal poor with [Fe/H]~$=$~$-1.07$ \citep{Limberg:2022, Cunningham:2024}.
For Antlia~II and Crater~II we adopt modeling estimates of $f_\text{bound}$~$=$~0.1 \citep{Sameie:2020} and 0.3 \citep{Sanders:2018} respectively\footnote{\citet{Ji:2021} argue against substantial mass loss for Antlia~II and Crater~II based on offset from the \citet{Kirby:2013} relation. We assign both objects to Class~2 where $f_\text{bound} \gtrsim 0.2$ is possible, such that the \citet{Sameie:2020} Antlia~II estimate of 0.1 is difficult to match but the \citet{Sanders:2018} Crater~II estimate of 0.3 is feasible.} and assume there is no shift in metallicity given the lack of a detected metallicity gradient \citep{Ji:2021}.

Even though these estimates are simple, they illustrate (based on independent modeling) how far individual satellites can travel in the mass-metallicity plane.
In addition, it is clear that their `total' mass and metallicity should not be compared to the mass-metallicity relation of satellites, but to the (unknown) intrinsic relation (as highlighted in Figure~\ref{fig:mass-feh-fbound}).
It is also intriguing that these reconstructions move their respective satellites near the lowest edge of the satellite mass-metallicity relation, suggesting that this may be near where the intrinsic relation lies.
These results imply a difference in the mass-metallicity relations between centrals and satellites, an effect already observed at higher masses \citep{Pasquali:2010}.
Simulators can also be more accepting of a mass-metallicity relation for isolated low-mass galaxies that is lower and steeper than what is observed in the Local Group, since it is the intrinsic relation that they seek to reproduce \citep[e.g.,][]{Bose:2025}.

Finally, when estimating the stellar mass of streams a common technique \citep[e.g.][]{Li:2022} is to measure the stream's metallicity and then adopt a mass-metallicity relation, often from \citet{Kirby:2013}.
We emulate this method for Auriga systems by computing the mean metallicity of unbound stars and converting this into a stellar mass using the evolved relation (solid line in Figure~\ref{fig:mass-feh-fbound}).
We find this underestimates the true total mass with $\log_{10}(M_\ast^\text{est} / M_\ast^\text{true}) = -0.34^{+0.38}_{-0.59}$ for systems with a progenitor at the present day and $\log_{10}(M_\ast^\text{est} / M_\ast^\text{true}) = -0.57^{+0.33}_{-0.34}$ for systems with no progenitor.
The latter is impacted by redshift evolution of the intrinsic relation (Figure~\ref{fig:mass-feh-t90}).
While this technique is an improvement over methods using star counts for only the chunk of stream that is detected, we recommend such estimates should be considered lower bounds on the total mass of the system.

\section{Concluding remarks} \label{sec:conclusions}

Accreted systems in Auriga follow established correlations between stellar mass, metallicity, and star formation history (Figure~\ref{fig:mass-feh-t90}) that, upon accretion and disruption, imprint a tidally evolved mass-metallicity relation that differs from the intrinsic one (Figure~\ref{fig:mass-feh-fbound}).
The scatter about this tidally evolved relation is similar to that observed for Local Group satellites (Figure~\ref{fig:mass-feh-observations}) and offset from the relation predicts which satellites have experienced substantial mass loss (Figure~\ref{fig:feh-diff-fbound}).
The mass-metallicity relation for Local Group satellites is not only compatible with tidal mass loss, but provides further evidence that a wealth of faint streams awaits discovery in Rubin, Euclid, and Roman data.

\section*{Acknowledgements}
We thank Astha Chaturvedi, Emma Dodd, Ivanna Escala, Azi Fattahi, Peter Ferguson, and Guilherme Limberg for enlightening discussions, as well as Jess Doppel and Louise Welsh for peer pressure to `just write the damn paper' and an anonymous referee for helpful suggestions.
AHR thanks Taylor Swift's \textit{1989 (Taylor's Version)} and Bring Me The Horizon's \textit{POST HUMAN: NeX GEn} for providing a soundscape condusive to writing this manuscript.
This research made extensive use of \href{https://arxiv.org/}{arXiv.org} and NASA's Astrophysics Data System for bibliographic information.

AHR was supported by a fellowship funded by the Wenner Gren Foundation, a Research Fellowship from the Royal Commission for the Exhibition of 1851, and by STFC through grant ST/T000244/1.
RB is supported by the UZH Postdoc Grant, grant no.~FK-23116 and the SNSF through the Ambizione Grant PZ00P2\_223532.
FF is supported by a UKRI Future Leaders Fellowship (grant no. MR/X033740/1).
FAG acknowledges support from the ANID BASAL project FB210003, from the ANID FONDECYT Regular grants 1251493 and from the HORIZON MSCA-2021-SE-01 Research and Innovation Programme under the Marie Sklodowska-Curie grant agreement number 101086388.
RJJG is supported by an STFC Ernest Rutherford Fellowship (ST/W003643/1).
FM acknowledges funding from the European Union - NextGenerationEU under the HPC project `National Centre for HPC, Big Data and Quantum Computing' (PNRR - M4C2 - I1.4 - CN00000013 – CUP J33C22001170001).
This work was supported by collaborative visits funded by the Cosmology and Astroparticle Student and Postdoc Exchange Network (CASPEN).
We acknowledge support from the DiRAC Institute in the Department of Astronomy at the University of Washington.
The DiRAC Institute is supported through generous gifts from the Charles and Lisa Simonyi Fund for Arts and Sciences, Janet and Lloyd Frink, and the Washington Research Foundation.

This work used the DiRAC@Durham facility managed by the Institute for Computational Cosmology on behalf of the STFC DiRAC HPC Facility (www.dirac.ac.uk).
The equipment was funded by BEIS capital funding via STFC capital grants ST/K00042X/1, ST/P002293/1, ST/R002371/1 and ST/S002502/1, Durham University and STFC operations grant ST/R000832/1.
DiRAC is part of the National e-Infrastructure.
This research used resources of the Argonne Leadership Computing Facility, a U.S. Department of Energy (DOE) Office of Science user facility at Argonne National Laboratory and is based on research supported by the U.S. DOE Office of Science-Advanced Scientific Computing Research Program, under Contract No. DE-AC02-06CH11357.

For the purpose of open access, the author has applied a Creative Commons Attribution (CC BY) licence to any Author Accepted Manuscript version arising from this submission.

\section*{Software}
This research made use of the Python programming language, along with many community-developed or maintained software packages including:
\begin{itemize}
    \item[(i)] \textsc{astropy} \citep{Astropy:2013, Astropy:2018, Astropy:2022},
    \item[(ii)] \textsc{cmasher} \citep{cmasher, cmasher1.8.0}
    \item[(iii)] \textsc{cython} \citep{cython}
    \item[(iv)] \textsc{h5py} \citep{h5py, h5py3.7.0}
    \item[(v)] \textsc{jupyter} \citep{ipython, jupyter}
    \item[(vi)] \textsc{matplotlib} \citep{matplotlib}
    \item[(vii)] \textsc{numpy} \citep{numpy}
    \item[(viii)] \textsc{pandas} \citep{pandas, pandas1.5.0}
    \item[(ix)] \textsc{scikit-learn} \citep{scikit-learn, scikit-learn-api, scikit-learn1.1.2}
    \item[(x)] \textsc{scipy} \citep{scipy, scipy1.9.1}.
\end{itemize}
We also thank the maintainers of the \textsc{arepo-snap-util} package.
Parts of the results in this work make use of the colormaps in the \textsc{cmasher} package.
Software citation information aggregated using \href{https://www.tomwagg.com/software-citation-station/}{The Software Citation Station} \citep{software-citation-station-paper, software-citation-station-zenodo}.

\section*{Data Availability}

Halo catalogs, merger trees, and particle data (Section \ref{sec:data}) for Auriga are publicly available \citep[detailed in the Auriga project data release;][]{Grand:2024} to download via the Globus platform\footnote{\href{https://wwwmpa.mpa-garching.mpg.de/auriga/data.html}{https://wwwmpa.mpa-garching.mpg.de/auriga/data}}.
The observational data used in this work is compiled in Table~\ref{tab:observations} largely thanks to the heroic efforts of Andrew Pace and Alan McConnachie \citep{McConnachie:2012, Pace:2025}.


\bibliographystyle{mnras}
\bibliography{main, software, lvdb}


\appendix

\section{Ignoring metallicity gradients} \label{app:ignore-gradients}

\begin{figure}
    \centering
    \includegraphics[width=1.0\linewidth]{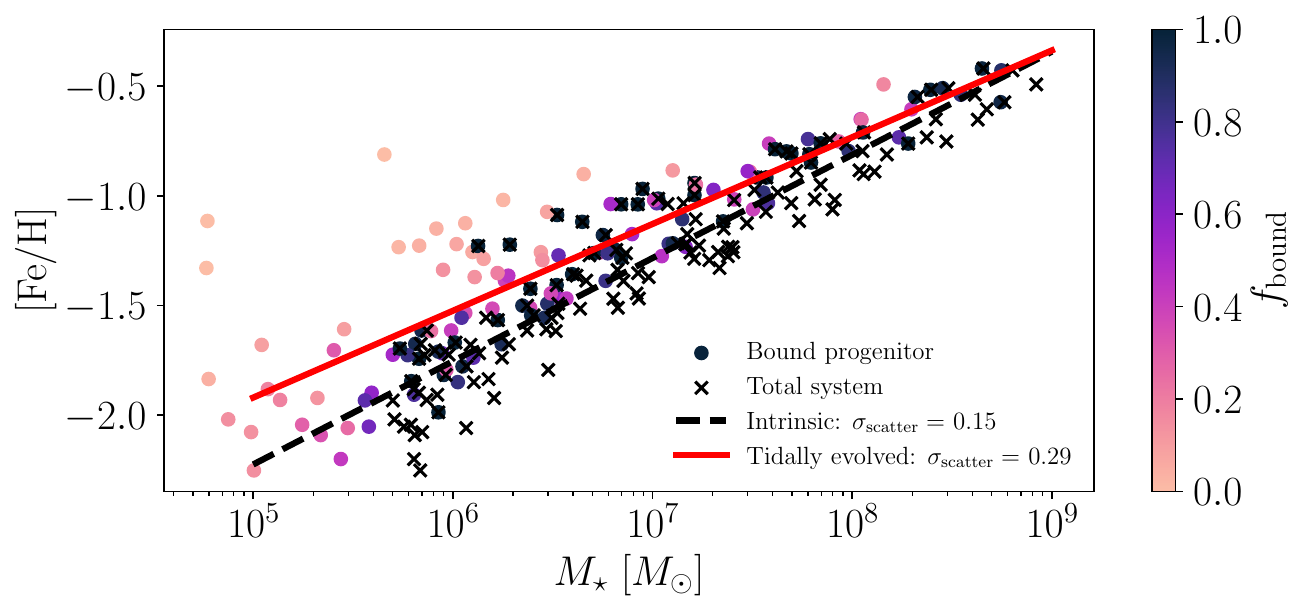}
    \caption{Similar to Figure~\ref{fig:mass-feh-fbound} but excluding the effect of radial metallicity gradients.}
    \label{fig:mass-feh-fbound-nogradient}
\end{figure}

Many of the key results in this analysis for Auriga are driven simultaneously by both tidal mass loss and by initial negative radial metallicity gradients.
Due to these gradients, low metallicity stars are preferentially lost from the outskirts, raising the mean metallicity of the bound progenitor.
While we prefer to present results that include both effects, it can be interesting to see if our results qualitatively hold when ignoring the effect of radial gradients.

In Figure~\ref{fig:mass-feh-fbound-nogradient} we show a version of Figure~\ref{fig:mass-feh-fbound} that ignores the metallicity evolution of the bound progenitor by adopting the metallicity of the total system.
In other words, individual points move exclusively to lower stellar mass.
We still find that the mean relation moves away from the intrinsic one, the scatter about the relation increases, and that individual points `invert' their relationship to the relation.

\section{Disruption prediction for observed satellites} \label{app:individual-sats}

\begin{table}
    \centering
    \begin{tabular}{cccccc}
        \hline
        Class & Definition & $f_\text{bound}^\text{Au}$ & $F^\text{Au}_\text{intact}$ & $F^\text{Au}_\text{<0.8}$ & $N^\text{obs}_\text{sat}$ \\
        & $\Delta \text{[Fe/H]} / \sigma \in$ \\
        \hline
        1 & $(-\infty, -1]$ & $0.92^{+0.07}_{-0.46}$ & 0.38 & 0.31 & 5 \\
        2 & $(-1, 0]$ & $0.81^{+0.18}_{-0.43}$ & 0.21 & 0.47 & 17 \\
        3 & $(0, 1]$ & $0.39^{+0.42}_{-0.26}$ & 0.07 & 0.83 & 12 \\
        4 & $(1, 2]$ & $0.06^{+0.05}_{-0.02}$ & 0.00 & 1.00 & 3 \\
        5 & $(2, \infty)$ & $0.00^{+0.01}_{-0.00}$ & 0.00 & 1.00 & 2 \\
        \hline
    \end{tabular}
    \caption{Properties of the different broad categories of disruption assigned to individual satellites.
    We provide the Class number, boundary definitions (see Section~\ref{sec:local-group-sats} and Figure~\ref{fig:feh-diff-fbound}), the distribution of $f_\text{bound}$ for Auriga satellites, fraction of Auriga satellites that are intact \citepalias[$f_\text{bound} > 0.97$, see][]{Riley:2025, Shipp:2025}, fraction of Auriga satellites with $f_\text{bound} < 0.8$, and number of Milky Way and M31 satellites in this Class.
    }
    \label{tab:class-definitions}
\end{table}

In Section~\ref{sec:relations} we introduce a broad classification for how disrupted Auriga satellites are, based on their offset from the tidally evolved mass-metallicity relation $\Delta$[Fe/H] normalised by the scatter about the relation $\sigma_\text{scatter}$.
The exact boundaries are listed in Table~\ref{tab:class-definitions} and separate into the following qualitative descriptions:
\begin{itemize}
    \item Class 1: unlikely to have experienced any mass loss
    \item Class 2: mass loss down to $f_\text{bound} \sim 0.2$ is possible but disfavoured
    \item Class 3: substantial mass loss is favoured, with most satellites having $0.05 < f_\text{bound} < 0.8$
    \item Class 4: no intact satellites, $f_\text{bound} < 0.2$ strongly preferred
    \item Class 5: no intact satellites, only $f_\text{bound} < 0.05$
\end{itemize}
Table~\ref{tab:class-definitions} also lists quantitative metrics for how disrupted Auriga satellites are in each Class.
As shown in Figure~\ref{fig:feh-diff-fbound}, both the highest level of disruption that is possible and the fraction of satellites that are substantially disrupted increase with ascending Class order.
We caution that these predictions are less robust for systems that have recent star formation (see final paragraph of Section~\ref{sec:relations}).

\begin{table*}
    \begin{tabular}{lcrrcc}
\hline
Name & $\log_{10} (M_\star / \text{M$_\odot$})$ & [Fe/H] & $\Delta$[Fe/H] & Class & References \\
\hline
\multicolumn{6}{c}{Milky Way satellites} \\
\hline
Antlia II & 6.13 & $-$1.9 & $-$0.15 & 2 & [1,1] \\
Canes Venatici I & 5.73 & $-$1.91 & $-$0.02 & 2 & [2,3] \\
Carina & 6.01 & $-$1.8 & $-$0.0 & 2 & [2,3] \\
Crater II & 5.51 & $-$2.16 & $-$0.19 & 2 & [4,1] \\
Draco & 5.78 & $-$2.0 & $-$0.13 & 2 & [2,3] \\
Fornax & 7.59 & $-$1.07 & 0.18 & 3 & [2,3] \\
Leo I & 6.96 & $-$1.48 & $-$0.02 & 2 & [2,3] \\
Leo II & 6.12 & $-$1.68 & 0.08 & 3 & [2,3] \\
LMC & 9.46 & $-$0.5 & 0.1 & 3 & [5,6] \\
Sagittarius & 7.63 & $-$0.53 & 0.7 & 5 & [7,3] \\
Sculptor & 6.54 & $-$1.73 & $-$0.12 & 2 & [2,3] \\
Sextans & 5.72 & $-$1.97 & $-$0.08 & 2 & [2,3] \\
SMC & 8.95 & $-$1.0 & $-$0.22 & 2 & [5,8] \\
Ursa Minor & 5.78 & $-$2.13 & $-$0.26 & 1 & [2,9] \\
\hline
\multicolumn{6}{c}{M31 satellites} \\
\hline
Andromeda I & 6.77 & $-$1.51 & 0.02 & 3 & [10,11] \\
Andromeda II & 6.92 & $-$1.47 & 0.01 & 3 & [10,12] \\
Andromeda III & 6.03 & $-$1.75 & 0.04 & 3 & [10,11] \\
Andromeda V & 5.95 & $-$1.84 & $-$0.03 & 2 & [10,11] \\
Andromeda VI & 6.75 & $-$1.5 & 0.04 & 3 & [13,14] \\
Andromeda VII & 7.35 & $-$1.37 & $-$0.04 & 2 & [13,11] \\
Andromeda IX & 5.69 & $-$2.03 & $-$0.12 & 2 & [10,12] \\
Andromeda XIV & 5.69 & $-$2.23 & $-$0.33 & 1 & [10,12] \\
Andromeda XV & 5.58 & $-$1.43 & 0.51 & 5 & [10,12] \\
Andromeda XVII & 5.35 & $-$1.7 & 0.32 & 4 & [10,14] \\
Andromeda XVIII & 5.9 & $-$1.53 & 0.3 & 4 & [10,15] \\
Andromeda XIX & 6.25 & $-$1.5 & 0.21 & 3 & [10,16] \\
Andromeda XXI & 5.81 & $-$1.7 & 0.16 & 3 & [10,17] \\
Andromeda XXIII & 6.14 & $-$2.2 & $-$0.45 & 1 & [10,14] \\
Andromeda XXV & 5.87 & $-$1.9 & $-$0.06 & 2 & [10,18] \\
Andromeda XXVII & 5.39 & $-$2.1 & $-$0.09 & 2 & [19,14] \\
Andromeda XXVIII & 5.64 & $-$1.84 & 0.08 & 3 & [20,20] \\
Andromeda XXIX & 5.5 & $-$1.9 & 0.07 & 3 & [20,20] \\
Cassiopeia II & 5.33 & $-$1.7 & 0.33 & 4 & [10,14] \\
Cassiopeia III & 7.19 & $-$1.7 & $-$0.32 & 1 & [21,22] \\
Lacerta I & 6.72 & $-$2.0 & $-$0.45 & 1 & [21,22] \\
NGC 147 & 8.17 & $-$0.83 & 0.22 & 3 & [7,23] \\
NGC 185 & 8.18 & $-$1.12 & $-$0.08 & 2 & [7,23] \\
NGC 205 & 8.84 & $-$0.92 & $-$0.1 & 2 & [7,23] \\
Perseus I & 5.78 & $-$2.0 & $-$0.13 & 2 & [21,22] \\
\hline
\end{tabular}
    \caption{Observational data (compiled by \citealt{Pace:2025}) presented in Figure~\ref{fig:mass-feh-observations} along with the vertical offset from the mass-metallicity relation ($\Delta$[Fe/H]) and a Class indicating the potential level of (tidal) disruption experienced by the system (see text in Appendix~\ref{app:individual-sats} for details).
    We also list references for each measurement, which correspond to the following citations: (1) \citet{Ji:2021}
(2) \citet{Munoz2018ApJ...860...66M}
(3) \citet{Simon:2019}
(4) \citet{Torrealba2016MNRAS.459.2370T}
(5) \citet{deVaucouleurs1991rc3..book.....D}
(6) \citet{Carrera:2008}
(7) \citet{McConnachie:2012}
(8) \citet{Parisi:2010}
(9) \citet{Pace2020MNRAS.495.3022P}
(10) \citet{Martin2016ApJ...833..167M}
(11) \citet{Kirby:2020}
(12) \citet{Wojno2020ApJ...895...78W}
(13) \citet{McConnachie2006MNRAS.365.1263M}
(14) \citet{Collins2013ApJ...768..172C}
(15) \citet{Kvasova2024ApJ...972..180K}
(16) \citet{Cullinane2024ApJ...972..133C}
(17) \citet{Collins2021MNRAS.505.5686C}
(18) \citet{Charles2023MNRAS.521.3527C}
(19) \citet{Richardson2011ApJ...732...76R}
(20) \citet{Slater2015ApJ...806..230S}
(21) \citet{Rhode2023AJ....166..180R}
(22) \citet{Martin2014ApJ...793L..14M}
(23) \citet{Kirby:2013}.}
    \label{tab:observations}
\end{table*}

We then apply the same exercise to the observed Local Group satellites to predict how much tidal disruption they have experienced.
We begin with compilations of their stellar masses and mean metallicities (Figure~\ref{fig:mass-feh-observations}), then fit a relation in the same manner as done for the Auriga satellites.
Our fit takes the following form:
\begin{equation} \label{eqn:our-fit}
    \text{[Fe/H]} = -1.80 + 0.35 \log_\text{10} \left( \frac{M_\star}{10^6\text{~M$_\odot$}} \right)
\end{equation}
The equivalent for \citet{Kirby:2013}, taking the luminosity-metallicity relation reported in their Equation~3 and assuming $M/L_V = 2$, is
\begin{equation} \label{eqn:kirby13}
    \text{[Fe/H]} = -1.77 + 0.29 \log_\text{10} \left( \frac{M_\star}{10^6\text{~M$_\odot$}} \right)
\end{equation} 
We compute each satellite's offset from the relation (Equation~\ref{eqn:our-fit}) normalized by the reported scatter ($\sigma_\text{scatter} = 0.23$).
These are then directly compared to the Class boundary definitions in Table~\ref{tab:class-definitions}.
The stellar masses, metallicities, offsets from the relation in Equation~\ref{eqn:our-fit}, and resulting Class assignments are indicated in Table~\ref{tab:observations}.
We note that Local Group satellites with detected tidal tails are typically in Classes 3-5 \citep[Sagittarius, NGC~147, And~XIX, And~XXI;][]{Ibata:2001, Crnojevic:2014, Collins2020MNRAS.491.3496C, Collins2021MNRAS.505.5686C}, with some in Class 2 \citep[Antlia~II, Crater~II, NGC~205;][]{Geha:2006, Ji:2021, Limberg:2025, Vivas:2025}.

We note that there are three M31 satellites (M~32, IC~10, and LGS~3) brighter than $M_V$~=~$-7.7$ omitted from our analysis because they lack literature spectroscopic metallicities in the compilations we consider.
The available isochrone or RGB colour metallicities compiled by \citet{McConnachie:2012} suggest that IC~10 and LGS~3 are not candidates for heavy disruption under our framework (Class~1 or 2 depending on the adopted relation).
However, the available data for M~32 gives it a very high metallicity \citep[$\text{[Fe/H]} = -0.25$;][]{Grillmair:1996} for its stellar mass.
This would place M~32 in Class~5, comparable to Sagittarius \citep[see also][]{Escala:2025}.


\bsp	
\label{lastpage}
\end{document}